\documentclass[12pt]{article}
\usepackage{amsfonts, latexsym, amssymb}

\newcommand{\ba}{\begin{array}}
\newcommand{\ea}{\end{array}}
\newcommand{\be}{\begin{equation}}
\newcommand{\ee}{\end{equation}}
\newcommand{\bea}{\begin{eqnarray}}
\newcommand{\eea}{\end{eqnarray}}
\newcommand{\bean}{\begin{eqnarray*}}
\newcommand{\eean}{\end{eqnarray*}}
\newcommand{\id}{1 \hspace{-1.1mm} {\rm I}}
\newcommand{\Rset}{\mathbb R}
\newcommand{\phiop}{\mbox{\boldmath$\varphi$}}
\newcommand{\Journal}[4]{{\it #1} {\bf #2} ({#4}) {#3}}

\begin{document}
\thispagestyle{empty}

\hfill IHES-P/00/xx
\vspace{3.5cm}

\begin{center}
{\LARGE 
Characterizing Volume Forms}
\ \ \\ \ \\ \ \\
{\large Pierre Cartier} \\ [2mm]
Ecole Normale Sup\'{e}rieure\\ 
45 rue d'Ulm F-75005 Paris \\[6mm]
{\large Marcus Berg, C\'{e}cile DeWitt-Morette
and Alex Wurm}  \\ [2mm]
Department of Physics and Center for Relativity, \\
University of Texas, Austin, TX 78712 
\end{center}
\ \\ \ \\ 
\begin{center}
{\bf Abstract}
\ \\ 
Old and new results for characterizing volume forms
in functional integration.
\end{center}

\newpage

\section{The Wiener measure}
Defining  \index{volume forms} volume forms on infinite dimensional spaces is a key problem in the
theory of functional integration. \index{functional integration}
 The first volume form
 used in functional
integration has been the Wiener measure. \index{Wiener measure} From the
 several equivalent
definitions of the Wiener measure, we choose
one \cite{Bourbaki} which can easily be extended
for use in Feynman integrals. We recall 
the Cameron-Martin and Malliavin formulae because they are,
respectively, integrated and infinitesimal formulae for changes of
variable of integration which can be imposed on volume forms other 
than the Wiener measure.

\subsection{Definition}
The Wiener measure \index{Wiener measure} $\gamma$
on the space ${\mathbb W}$ of pointed continuous paths
$w$ on the time interval $T=[0,1]$
\[
w : [0,1] \longrightarrow {\mathbb R}\;  ,\qquad\qquad 
w(0) =0,
\]
can be characterized by the equation
\be
\int_{\mathbb W} d\gamma (w)\; \exp\left( - i \langle w',
w \rangle\right) = \exp \left( -\frac{1}{2}\int_T \int_T dw'(t)\;
dw'(t')\; \inf(t,t')\right)\label{a1}
\ee
where $w'$ is an element of the topological dual ${\mathbb W'}$ of
${\mathbb W}$, i.e. a bounded measure on the semi-open 
interval $T=]0,1]$
\be
\langle w',w \rangle = \int_T dw'(t)\; w(t)\; .
\label{a3}
\ee

\subsection{Cameron-Martin formula}
The Cameron-Martin formula \index{Cameron-Martin formula} can be written
\be
\int_{\mathbb W} d\gamma(w)\; F\!\left(w +\varphi\right) =
\int_{\mathbb W} d\gamma(w)\; J\!\left(\varphi,w\right)\; F(w)
\label{a4}
\ee
where $J\!\left(\varphi,w\right)$ is the Radon-Nikodym derivative
\be
J\!\left(\varphi,w\right) = \frac{d\gamma(w-\varphi)}{d\gamma(w)} \; .
\label{a5}
\ee
This formally obvious expression for $J\!\left(\varphi,w\right)$ has 
a not-so-obvious explicit expression; when $\varphi\in L^{2,1}$, i.e.
when $\dot{\varphi}(t)$ is square integrable, and $\varphi(0)=0$, 
then\footnote{The boundary condition $\varphi(0)=0$ is required in order
that $w+\varphi$ belong to ${\mathbb W}$ when $w$ does, since $w(0)=0$
for every element $w$ of ${\mathbb W}$.}
\be
J\!\left(\varphi,w\right) = \exp\left(-\frac{1}{2}\int_T dt\; 
\dot{\varphi}(t)^2 +\int_T dw(t)\; 
\dot{\varphi}(t)\right).\label{a6}
\ee
The meaning of the second term is subtle since $w$ is not of bounded variation.
If $\varphi\in C^2(0,1)$ with the boundary conditions $\varphi(0) =
\dot{\varphi}(1)=0$, the second term can be integrated by parts --- giving
 $-\int_T dt\; w(t)\,\ddot{\varphi}(t)$ ---  and then can be
extended by continuity in the $L^{2,1}$ norm for $\varphi$.
\ \\ \ \\
A heuristic proof of the Cameron-Martin formula \index{Cameron-Martin formula} 
makes the explicit expression (\ref{a6}) ``obvious''.
Let us write formally
\be
d\gamma(w) = {\mathcal D}w\; \exp\left(-\pi Q(w)\right)
\label{a7}
\ee
where ${\mathcal D}$ is a translation-invariant
 symbol\footnote{The symbol ${\cal D}$ is often used in physics 
where ${\cal D}w:=\prod_t dw(t)$. Here it is defined by (\ref{a9}),
which in finite dimensions reduces to (\ref{finite2}).}
\be
{\mathcal D}\!\left(w+\varphi\right) = {\mathcal D}w
\label{a8}
\ee
and where $Q$ follows from the definition (\ref{a1}) of the Wiener
measure:
\be
\int_{\mathbb W} {\mathcal D}w\; \exp\left(-\pi Q(w)\right)\; 
\exp\left(-2\pi i \langle w',w\rangle\right) =
\exp\left( -\pi W\!\left(w'\right)\right)\label{a9}
\ee
with
\be
W\!\left(w'\right) = 2\pi \int_T\int_T dw'(t)\; dw'(t') 
\; \inf(t,t')\; .
\label{a10}
\ee
By analogy with the finite dimensional case (\ref{finite}), the
quadratic form  $Q$
 on ${\mathbb W}$ is required to be the inverse
of $W$ on ${\mathbb W'}$ in the following sense. Represent $W$ and $Q$ as
\be
W(w') = \langle w', Gw'\rangle\qquad\mbox{and}\qquad Q(w) = 
\langle Dw,w\rangle ;\label{a11}
\ee
then $Q$ is said to be the inverse of $W$ if
\be
D G = \id \; .
\label{a12}
\ee
It follows from (\ref{a10}) and (\ref{a11}) that
\be
Gw'(t) = 2\pi \int_T dw'(t')\; \inf (t,t').
\ee
It follows from (\ref{a12}) and (\ref{a11}) that
\be
Q(w) = \frac{1}{2\pi} \int_T dt\; \left(\frac{dw(t)}{dt}\right)^2
= \frac{1}{2\pi} \int_T \frac{\left( dw(t)\right)^2}{dt} \; .
\label{a13}
\ee
The Cameron-Martin formula \index{Cameron-Martin formula} (\ref{a4}) is now the obvious statement
\be
\int_{\mathbb W} {\mathcal D}w\; \exp\left(-\pi Q(w)\right)\; 
F\!\left( w+\varphi\right) =
\int_{\mathbb W} {\mathcal D}w\; \exp\left(-\pi Q(w-\varphi)\right)\; 
F (w)\label{a14}
\ee
that is $J(\varphi,w) = \exp \pi\left( Q(w)-Q(w-\varphi)\right)$. We
calculate
\bea
Q\! \left( w-\varphi\right) & = & \frac{1}{2\pi} \int_T dt\; \left( 
\frac{d}{dt} (w-\varphi) (t)\right)^2\nonumber\\
 & = & \frac{1}{2\pi} \left( Q(w) + \int_T dt\; \dot{\varphi}(t)^2
 - 2 \int_T dw(t)\; \dot{\varphi}(t)
\right)\; ; 
\label{a15}
\eea
formula (\ref{a6}) follows immediately and this completes the heuristic
 demonstration. 

\subsection{An analogy with the Dobrushin-Lanford-Ruelle
characterization of Gibbs states}

What is missing in the heuristic proof of the \index{Cameron-Martin formula}
Cameron-Martin formula 
to be rigorous? The difficulty is that the Brownian trajectories $w$
are so rough that $Q(w)$ is infinite if calculated as
the limit of the Riemann sums $\sum_{i=1}^N \left( \Delta\, w_i\right)^2
/\Delta\, t_i$ where
\[
 0= t_0<t_1<\ldots <t_N =1,\quad \Delta t_i = t_i-t_{i-1}\quad
\mbox{and}\quad \Delta\, w_i = w(t_i)-w(t_{i-1}).
\]
However, in the \index{Cameron-Martin formula} Cameron-Martin formula we need only the difference
$Q(w)-Q(w-\varphi)$. The infinite part drops out in the difference
provided $\varphi$ is smooth enough; e.g. if $\varphi$ is of
class $C^2$ on $T=[0,1]$.\\
A similar situation occurs in statistical mechanics in the case of infinite
volumes. For a configuration $w$, the formal Hamiltonian $H(w)$ may be
infinite. But if a configuration $w'$ is obtained by a local modification
of $w$ --- by changing the states in finitely many sites --- then
$H(w)-H(w')$ is finite. This is the strategy underlying the
 Dobrushin-Lanford-Ruelle characterization of Gibbs states.\cite{Galla}

\subsection{Malliavin formula}
 According to the \index{Malliavin formula} Malliavin formula,
\be
\int_{\mathbb W} d\gamma(w)\; D_{\varphi} F(w) = \int_{\mathbb W}
d\gamma(w)\; A_{\varphi}(w) F(w)\label{a16}
\ee
where $D_{\varphi}F$ is the Gateaux differential of $F$ in the
$\varphi$-direction,
\be
D_{\varphi} F(w) = \lim_{\epsilon=0} \frac{1}{\epsilon} \left(
F(w+\epsilon\varphi) - F(w)\right) = \int_T dt\;
\frac{\delta F(w)}{\delta w(t)}\; \varphi(t)\; ,\label{a17}
\ee
and where
\bea
A_{\varphi}(w) & := & \int_T dw(t)\; \dot{\varphi}(t)
\nonumber\\
& = & - \int_T dt\; w(t)\; \ddot{\varphi}(t)\label{a18} 
\eea
for $\varphi\in C^2(0,1)$, with $\varphi(0)=\dot{\varphi}(1)=0$.
Again, the Malliavin formula \index{Malliavin formula} is ``obvious'' if we use the formal
expression (\ref{a7}) and integrate the left hand side of (\ref{a16})
by parts,
\be
\int_{\mathbb W} {\mathcal D}w\; \exp\left(-\pi Q(w)\right)\; 
D_{\varphi} F(w) = -\int_{\mathbb W}{\mathcal D}w\; D_{\varphi}
\!\left(\exp\left(-\pi Q(w)\right)\right)\;  F(w)
\; .\label{a19}
\ee
The \index{Cameron-Martin formula} Cameron-Martin formula  and its infinitesimal form, the
 \index{Malliavin formula} Malliavin formula, pave the way for defining a
 formal
 translation-invariant
symbol ``${\mathcal D}$''.
In this paper, we propose an infinitesimal 
characterization of ${\cal D}$;
namely, given an arbitrary functional $U$ integrable by ${\mathcal D}w$,
 the translation invariance of ${\cal D}$ can be expressed by
integrating by parts
\be
\int_{\mathbb W} {\cal D}w {\delta U \over \delta w(t)} = -
 \int_{\mathbb W} 
\frac{\delta}{\delta w(t)} {\cal D}w\cdot U = 0 \qquad\qquad \forall\, U \; .
\ee

\subsection{Some lessons from the Malliavin formula}

\begin{itemize}
\item Let us write formally
\be
d\gamma(w)=\mu(w)\, {\cal D}w
\ee
where $\mu(w)$, often called ``measure'' in physics, is not necessarily
$\exp\left(-\pi Q(w)\right)$.
\ \\ \ \\
Malliavin's formula reads
\be
\int_{\mathbb W} \mu(w)\,{\cal D}w\left( D_{\varphi}
 F(w) - A_{\varphi}(w)\, F(w)\right) = 0
\ee
which, by a formal integration by parts, becomes
\be
\int_{\mathbb W} {\cal D} w \left( D_{\varphi} \mu(w) + A_{\varphi}(w)\,
\mu(w)\right)\,
F(w) = 0 \; .
\ee
Since $F(w)$ is arbitrary, Malliavin's formula is equivalent to
\be
D_{\varphi} \mu(w) + A_{\varphi}(w)\,\mu(w) =0\label{twotwo}
\ee
for all $\varphi$ sufficiently regular, e.g. $\varphi\in C^2(0,1)$.
The ``measure'' $\mu(w)$ is, modulo a multiplicative constant, characterized
 by (\ref{twotwo}).
If $\mu(w) = \exp\left(-\pi Q(w)\right)$ as before, then
\be
D_{\varphi}Q(w) = \frac{1}{\pi} A_{\varphi}(w).
\ee
According to (\ref{a18}), $A_{\varphi}(w)$ is bilinear in $\varphi$ and $w$; 
hence $Q$ is quadratic in $w$ and we recover formula (\ref{a13}).
\item The above remark is a heuristic proof that the Wiener measure \index{Wiener measure}
 $\gamma$
is characterized by Malliavin's formula. The proof can be made rigorous
by choosing in (\ref{a16})
\[
F(w) = \exp\left(-2\pi i \langle w',w\rangle\right)\; , 
\qquad w'\in{\mathbb W'} \; .
\]

\item Conversely the \index{Cameron-Martin formula} Cameron-Martin formula provides a rigorous proof
of the \index{Malliavin formula} Malliavin formula (\ref{a16}): replace $\varphi$ by $\epsilon
\varphi$ in (\ref{a4}) and take the derivative of both sides with
respect to $\epsilon$. At $\epsilon=0$, one checks that
\[
\frac{d}{d\epsilon} J\!\left( \epsilon\varphi ,w\right)
\Bigg|_{\epsilon=0} = A_{\varphi}(w) \; .
\]
\item The Malliavin formula can be used for realizing creation and
annihilation operators on bosonic Fock spaces,\cite{Cartier} thanks to
the Wiener chaos isomorphism:
Let ${\cal H}:=L^{2,1}(T)$ be a one-particle (real) Hilbert space with
 scalar product $(\varphi_1|\varphi_2) = 
\int_T dt\; \dot{\varphi}_1(t)\,\dot{\varphi}_2(t)$. Let ${\cal F}
\!\left( {\cal H}\right)$ be the \index{Fock space} Fock space with vacuum $\Omega$. The
{\it Wiener chaos} \index{Wiener chaos} is an isomorphism (see e.g.\ \cite{Janson})
\be
L^2\!\left({\mathbb W},d\gamma\right) \simeq {\cal F}\!\left( {\cal H}\right).
\label{twofour}
\ee
With
\[
A_{\varphi}(w):=\int_T dw(t)\; \dot{\varphi}(t) ,
\]
one obtains
\be
\int_{\mathbb W} d\gamma(w)\; A_{\varphi_1}(w)\; A_{\varphi_2}(w) =
\int_T dt\; \dot{\varphi}_1(t)\; \dot{\varphi}_2(t) .\label{twoseven}
\ee
It is enough to consider the case $\varphi_1 =\varphi_2 =\varphi$.
This fundamental formula can then be established by a linear change of variable
of integration in ${\mathbb W}$,
\[
A_{\varphi} : w \longmapsto \int_T dw(t)\; \dot{\varphi}(t) .
\] 
Eq.~(\ref{twoseven}) says that the map from ${\cal H}$ to ${\cal F}$,
\bean
{\cal H} := L^{2,1}(T) & \longrightarrow & {\cal F} := L^2 \left(
{\mathbb W},d\gamma\right)\\
\mbox{ by}\qquad \qquad\qquad \varphi & \longmapsto & A_{\varphi}
\eean
is an isometry.\\
The Malliavin formula \index{Malliavin formula} (\ref{a16}), with
$\varphi =\varphi_1$ and $F=A_{\varphi_2}\,$, gives another proof
 of (\ref{twoseven}) 
\be
\left( \varphi_1 | \varphi_2\right)= \int_{\mathbb W} d\gamma(w)\; 
D_{\varphi_1}\! \left(A_{\varphi_2}(w)\right) =
\int_{\mathbb W} d\gamma(w)\; A_{\varphi_1}(w)\; A_{\varphi_2}(w)
\ee
and the following commutation relations are obvious
\bean
\left[ D_{\varphi_1}\; ,\; D_{\varphi_2}\right] & = & 0\\
\left[ A_{\varphi_1}\; ,\; A_{\varphi_2}\right] & = & 0\\
\left[ D_{\varphi_1}\; , \; A_{\varphi_2}\right] & = & \left(\varphi_1 |\varphi_2
\right) .
\eean
Therefore
\be
a^{\dagger}(\varphi) := A_{\varphi} - D_{\varphi} \qquad\qquad\mbox{and}
\qquad\qquad a(\varphi) := D_{\varphi}
\ee
obey the bosonic commutation relation of creation and annihilation
operators on ${\cal F}$, respectively, 
\[
\left[ a\!\left(\varphi_1\right)\; ,\; a^{\dagger}\!\left(\varphi_2\right)
\right] = \left( \varphi_1 | \varphi_2\right)\cdot \id ,
\]
other commutators vanishing. It can be proved that $a$ and $a^{\dagger}$ are
adjoint in the Hilbert space $L^2\!\left( {\mathbb W},d \gamma\right)$ by
integrating by parts the Malliavin formula \index{Malliavin formula} (\ref{a16}) with $F= F_1\, F_2$
\be
\int_{\mathbb W} d\gamma\; D_{\varphi} F_1 \cdot F_2 = \int_{\mathbb W}
d\gamma\; F_1 \cdot \left( A_{\varphi} F_2 - D_{\varphi} F_2\right) .
\ee
The vacuum $\Omega\in {\cal F}$ is the constant function equal to $1$. If 
a functional $F$ of the Brownian path $w$ acts on ${\cal F}$ by multiplication,
i.e.
\[
F(w) : \Psi (w) \longmapsto F(w)\, \Psi(w)
\]
then we derive the tautology
\be
\langle \Omega | F| \Omega\rangle = \int_{\mathbb W} d\gamma(w) \; F(w) .
\label{threeone}
\ee
{\it The \index{Wiener measure} Wiener measure is therefore the spectral measure corresponding to the
vacuum state $\Omega$}. The vacuum is characterized by
\[
a(\varphi)\, \Omega = D_{\varphi} \Omega = 0\qquad ,\qquad \forall\,\varphi\,
;
\]
alternatively
\[
\frac{\delta\,\Omega (w)}{\delta\, w(t)} = 0 \qquad \mbox{for every}\qquad
t\in T .
\]
With the notation of (\ref{threeone}) we write
\[
0 = \langle F | a(\varphi)\, \Omega\rangle = \langle a^{\dagger}(\varphi)\, F |
\Omega\rangle .
\]
Hence the vacuum $\Omega$ is, up to a scalar, the unique state orthogonal to
all $a^{\dagger}(\varphi)F$, that is to the functions $D_{\varphi}F
-A_{\varphi}\cdot F$. This gives another interpretation to Malliavin's 
formula (\ref{a16}).
\ \\ \ \\
To show that $\left\{ L^2\!\left({\mathbb W},d\gamma\right),\, \Omega ,\,
a(\varphi),\, a^{\dagger}(\varphi)\right\}$ is indeed a model of Fock space, it remains to check that the vectors, $a^{\dagger}\!\left(\varphi_1\right)\ldots
a^{\dagger}\!\left(\varphi_n\right) \Omega\,$, make a total system for
$L^2\!\left({\mathbb W},d\gamma\right)$ --- i.e. that the finite linear
combinations of $A_{\varphi_1}\ldots A_{\varphi_n}$ are dense in $L^2\!\left(
{\mathbb W}, d\gamma \right)$. The Wiener chaos \index{Wiener chaos} follows from the general theory
 of \index{Fock space} Fock space. The general theory includes not only the symmetric space
considered here, but also the antisymmetric Fock space which we have not yet 
considered.
\end{itemize}

\subsection{Feynman volume form (see refs 3,4)}
 The Fourier transform of the Wiener measure \index{Wiener measure} (\ref{a1}) or (\ref{a9}) suggests
a characterization of the \index{Feynman volume form} Feynman volume form by its Fourier transform.
Let $s\in \{ 1,i\}$, then we can define ${\mathcal D}x$ by
\be
\int_{\mathbb X} {\mathcal D}x\; \exp\left(-\frac{\pi}{s} Q(x)\right)\;
\exp\left(-2\pi i \langle x',x\rangle\right) = \exp\left(
-s \pi W(x')\right)
\label{defDx}
\ee
where ${\mathbb X}$ is the space of paths $x$ and the quadratic form
\bean
Q(x) > 0 & \mbox{ for } & s=1 \\
\mbox{imaginary part of } Q(x) > 0 & \mbox{ for } & s=i \; .
\eean
The case $s=1$ corresponds to the Wiener measure while the case $s=i$
corresponds to the Feynman sum over paths in quantum mechanics.
Everything said before can be repeated with obvious changes, e.g. the
\index{Malliavin formula} Malliavin formula.

\section{Volume forms in quantum field theory; \\Schwinger's dynamical
principle}

The functional integral representation of the Schwinger dynamical
principle has led Bryce DeWitt to the introduction of a ubiquitous
volume form in quantum field theory. 
According to Schwinger, the
variation of the probability amplitude for a transition $\langle {\rm
out} |
{\rm in} \rangle$ is given by the variation of the action ${\bf S}$ 
of the system\footnote{We use boldface for operators on Fock space.}:
\be
\delta \, \langle {\rm out} \, |\, 
{\rm in} \rangle = {i \over \hbar}\,  \langle {\rm out} | \,
\delta {\bf S} \, | {\rm in} \rangle 
\label{Schwinger}
\ee
where ${\bf S}$ is a functional of the field operators, which are
globally designated
by ${\mathbf \phiop}$.

\subsection{Evolution equations for the field operators $\phiop$}

 The \index{Schwinger variational principle}Schwinger variational principle
 gives evolution
equations for the field operators $\phiop$ different from the
classical Euler-Lagrange equations
\be
{\delta S \over \delta \varphi} = 0 \; .
\ee
The Schwinger-Dyson
equations give the quantum evolution of polynomials of fields $F(\phiop)$
for a system with classical action $S$ by the expectation value of a time
ordered operator,
\be
\langle {\rm vac} | \, T\left( {i \over \hbar} {\delta S \over \delta
\phiop} F(\phiop) + {\delta F \over \delta \phiop} \right)|{\rm vac}
\rangle = 0 \; .
\label{Schwinger-Dyson}
\ee
(proof of eq.\ (\ref{Schwinger-Dyson}) and some of its applications
can be found e.g.\ in
the textbook by Peskin and Schroeder \cite{Peskin},
section 9.6). 

\subsection{Functional integral solution of the Schwinger principle}

To exploit the \index{Schwinger variational principle} Schwinger variational
 principle (\ref{Schwinger}), one
varies an external source $J$ added to the original action $S$. The
new action is 
\[
S+\langle J,\phiop 
\rangle
\]
and the principle (\ref{Schwinger}) now reads
\[
{\hbar \over i}{\delta \over \delta J} 
\langle {\rm out} \, |\,  {\rm in} \rangle  =
\langle {\rm out} |\, \phiop \,  |  {\rm in} \rangle .
\]
Bryce DeWitt has constructed the following functional integral
solution of this equation (for details, see pp.\ 4160-4164 and related
references in \cite{JMP00}):
\be
\langle {\rm out} \, |\,  {\rm in} \rangle
= {\cal N} \int_{\Phi{\rm (in,out)}} \mu(\varphi)\,  {\cal D}\varphi\,
\exp\left({i \over \hbar}(S(\varphi)+\langle J, \varphi \rangle)\right)
\label{inout}
\ee
where 
\begin{itemize}
\item
${\cal N}$ is a normalization constant,
\item the domain of integration is defined by the {\it in} and
{\it out} states,
\item ${\cal D} \varphi$ is invariant under translations,
\item $\mu(\varphi)$ is, to leading order, given by the advanced Green's
function $G^+$:
\be
\mu(\varphi) = |\mbox{sdet } G^+(\varphi)|^{-1/2} + \ldots
\label{mu}
\ee
where ``sdet'' is the superdeterminant. 
In (\ref{mu}) the advanced Green's function $G^+$ is the unique inverse of
the leading nonconstant term $S''$ of the expansion  of $S$ restricted to
 its domain of integration $\Phi{\rm (in, out)}$:
\be
S(\varphi_0+\delta\phiop) = S(\varphi_0)+{1 \over 2}
\, S''(\varphi_0) \delta\phiop \,\delta\phiop + \ldots\; , \qquad
 S'(\varphi_0)=0.
\label{Greens}
\ee
\end{itemize}
Equations (\ref{defDx}) and (\ref{inout}) define volume forms
$\mu(\varphi){\mathcal D}\varphi$ and ${\mathcal D}x$ respectively. In
both cases ${\mathcal D}$ is a translation-invariant symbol. For
comparing the structure of the two volume forms we write the finite
dimensional version of (\ref{defDx}) with $s=1$:
\be
\int_{\Rset^D} {\mathcal D}x\; \exp(-\pi Q_{\alpha \beta} x^{\alpha}
x^{\beta}) \exp(-2 \pi i x'_{\alpha} x^{\alpha})  
= \exp(-\pi W^{\alpha \beta} x'_{\alpha} x'_{\beta})
\label{finite}
\ee
where
\bea
Q_{\alpha \beta}W^{\beta \gamma}&=&\delta^{\gamma}_{\alpha} \\
{\mathcal D}x &=& (\det Q_{\alpha \beta})^{1/2}\, dx^1 \ldots dx^D 
\nonumber\\
&=& (\det W^{\alpha \beta})^{-1/2}\, dx^1 \ldots dx^D \; .\label{finite2}
\eea
 Hence $G^+$ is to field theory what $W$ is to a
Gaussian on $\Rset^D$, that is, the covariance matrix:
\[
W^{\lambda \mu} = 2\pi \int_{{\mathbb R}^D} {\cal D}x\; \exp \left( -
\pi Q_{\alpha\beta} x^{\alpha} x^{\beta}\right) \; x^{\lambda} x^{\mu}.
\]
\ \\ \ \\
In (\ref{inout}) the term $(2\pi i /h )\langle J, \varphi
\rangle$ corresponds to $-2\pi i \langle x', x \rangle $ in
(\ref{defDx}); they both mean that the l.h.s.\ is a Fourier transform.
\ \\ \ \\
In finite dimensions it is easy to write explicitly ${\mathcal D}x$;
in infinite dimensions it has meaning only in the context of the
integral (\ref{defDx}). However, this implicit definition is
sufficient for computing functional integrals.\cite{JMP95,JMP00} It
is also easy to generalize it to cases other than Gaussians. 
\ \\ \ \\
Using a generalized formulation of ${\mathcal D}x$, or using
$\mu(\varphi){\mathcal D}\varphi$ obtained from the solution of the
\index{Schwinger variational principle} Schwinger variational principle is
 a matter of choice---often
dictated by context. 
Once $\mu(\varphi)$, or at least its leading term, is defined, the
following equation 
\[
\langle {\rm out} \, |\,  T(F(\phiop)) \, | \, {\rm in} \rangle
= {\mathcal N} \int \mu(\varphi) {\mathcal D}\varphi \, 
F(\varphi) \exp({i \over \hbar}\, S(\varphi))
\]
can be exploited in a variety of cases; i.e.\ in cases where
$F(\varphi)$ is not simply ${\rm exp}(i\langle J,\varphi \rangle
/ \hbar)$. 

\section{Volume forms in differential geometry}
We shall use differential geometry for defining 
\index{volume forms}volume forms on finite dimensional Riemannian and symplectic
manifolds in a formulation which paves the way for the infinite
dimensional case. Knowledgeable reader for whom using the infinite limit
of a finite volume element is---rightly---anathema, please bear with
us. Finite dimensional volume elements are useful in the
following situations.
\begin{itemize}
\item
A rule of thumb. A statement which is independent of the dimension of
the space of interest has a chance to generalize to infinite
dimensional spaces; for example a Gaussian on $\Rset^D$ defined by
(\ref{finite})  generalizes easily to (\ref{a9}). 
\item
Infinite dimensional spaces defined by a projective system of finite
dimensional spaces. This strategy was used in defining Feynman volume forms
by their Fourier transforms. \cite{Cecile72,Cecile79} 
\item
Differential calculus on Banach spaces, and differential geometry on
Banach manifolds. They are natural generalizations of their finite
dimensional counterparts. For this reason we propose a formula
which defines volume elements by their Lie derivatives.
\end{itemize}
Let ${\mathcal L}_X$ be the \index{Lie derivative} Lie derivative
 with respect to a vector field $X$ on a $D$-dimensional manifold
 $M^D$, either a (pseudo-)Riemannian manifold $(M^D,g)$
or a symplectic manifold $(M^{2N},\Omega)$;
the volume forms are, respectively,
\be
\omega_g(x) = |\det g_{\alpha \beta}(x)|^{1/2} dx^1 \wedge 
\ldots \wedge dx^D \qquad \mbox{ on } (M^D,g) 
\label{omegaR}
\ee
and
\be
\omega_{\Omega}(x) 
= \frac{1}{N!}\Omega \wedge \ldots \wedge \Omega \; \mbox{ (N factors) }
\qquad \mbox{ on } (M^{2N},\Omega).
\ee
In canonical coordinates $(p,q)$,
\[
\Omega = \sum_{\alpha} dp_{\alpha} \wedge dq^{\alpha} 
\]
and
\[
\omega_{\Omega} = dp_1 \wedge dq^1 \wedge \ldots \wedge
dp_N \wedge dq^N \; .
\]
Surprisingly $\omega_g$ and 
$\omega_{\Omega}$ satisfy equations of the
same structure:
\bea
{\cal L}_X \omega_g & = & 
{\scriptstyle {1 \over 2}}
{\rm Tr}\left(g^{-1}{\cal L}_X g\right) \omega_g 
\label{Divg}\\
{\cal L}_X \omega_{\Omega} & = &
{\scriptstyle {1 \over 2}} 
{\rm Tr}\left(\Omega^{-1}{\cal L}_X \Omega
\right) \omega_{\Omega}  \label{DivO} 
\eea
Riemannian and symplectic geometry are notoriously different
(see e.g. McDuff\cite{McDuff}) and the analogies between
them are not superficial. For instance,
with Riemannian geometry on the left and symplectic geometry on the right
\be
\begin{array}{ccc}
\int ds &\qquad& \int \Omega\\[2mm]
\mbox{geodesics} && \mbox{minimal surfaces}\\[2mm]
{\cal L}_X g = 0 \mbox{ defines}
&& {\cal L}_X \Omega =0 \mbox{ defines} \\ 
\mbox{Killing vector fields.} 
&& \mbox{Hamiltonian vector fields.}
\end{array}\label{killing}
\ee
Killing vector fields are few, Hamiltonian vector fields are many.

\subsection{The general case}
Before proving (\ref{Divg}), (\ref{DivO})
we consider the more general equation
\be
{\cal L}_X \omega = D(X)\cdot\omega
\ee
or its integrated formulation
\be
\int_M \left({\cal L}_X F\right)\;\omega = - \int_M F\; {\cal L}_X \omega
= - \int_M F\; D(X)\cdot \omega
\ee
where $\omega$ is a top form (a $D$-dimensional form on $M^D$) and
$D(X)$ is a function on $M$ depending on the vector field $X$ on $M$.
\begin{itemize}
\item Properties of $D(X)$ dictated by properties of ${\cal L}_X$:
\be
{\mathcal L}_{[X,Y]} = {\mathcal L}_X {\mathcal L}_Y
- {\mathcal L}_Y {\mathcal L}_X \; \Leftrightarrow \; 
D([X,Y]) = X\left( D(Y)\right) - Y\left( D(X)\right)
\ee
On a top form:
\be
{\mathcal L}_{fX} = f{\mathcal L}_X + X(f) \;
\Leftrightarrow \; 
D(fX) = f D(X)
+X(f) 
\ee
Proof: On a top form $\omega$, the Cartan formula
${\mathcal L}_X = d\, i_X + i_X\, d $ yields
\[
{\mathcal L}_X \omega = d \, i_X \omega \; ,
\]
and since $i_{fX} \omega = i_X \left(f\omega\right)$, we have
 ${\mathcal L}_{fX} \omega = d\, i_X
\left( f\omega\right) = {\cal L}_X\left(f\omega\right)$. $\blacksquare$
\item In coordinates,
\be
\omega_{\mu}(x) = \mu(x)\; dx^1\wedge\ldots\wedge dx^D = \mu(x)\, d^D x
\; .
\label{omega}
\ee
By the Leibniz rule,
\be
{\mathcal L}_X (\mu \, d^Dx) = {\mathcal L}_X (\mu)\, d^Dx+
\mu \, {\mathcal L}_X\!\left( d^Dx\right) \; .
\label{Leibniz}
\ee
Because $d^Dx$ is a top form on $M^D$, 
\be
{\mathcal L}_X\!\left( d^Dx\right) = d(i_X d^Dx) = {X^{\alpha}}_{,\, \alpha} \; d^Dx \; .
\label{Cartan}
\ee
Finally, combining (\ref{omega}), (\ref{Leibniz}) and (\ref{Cartan}),
\bea
{\mathcal L}_X \, \omega_{\mu} &=& (X^{\alpha} \mu_{,\alpha} + 
\mu \, {X^{\alpha}}_{,\alpha}) \, d^Dx  \nonumber \\
&=& D(X) \cdot \omega_{\mu} \label{fourthree}
\eea
with 
\bea
D(X) &=& \left(X^{\alpha} \mu_{,\alpha} + 
\mu \, {X^{\alpha}}_{,\alpha}\right)\,\mu^{-1}  \nonumber \\
&=& {X^{\alpha}}_{,\alpha}+ X^{\alpha}(\log |\mu|)_{,\alpha}
\; . \label{fourfour}
\eea
\end{itemize}

\subsection{The Riemannian case $(M,g)$}

Let $\omega_g(x) = \mu(x) d^Dx$. We shall show that the basic equation
(\ref{Divg}) is satisfied if and only if $\mu(x) = {\rm const}\; 
 |\det g(x) \, |^{1/2}$. Indeed
\[
({\mathcal L}_X g)_{\alpha \beta} = X^{\gamma} g_{\alpha \beta,\gamma}
+ g_{\gamma \beta}{X^{\gamma}}_{,\alpha} + 
g_{\alpha \gamma}{X^{\gamma}}_{,\beta}
\]
and
\be
{\rm Tr}\left( g^{-1} {\cal L}_X g\right) = g^{\beta\alpha} X^{\gamma}
g_{\alpha\beta,\gamma} + 2 X^{\alpha}_{\; ,\alpha}\label{foursix}
\ee
and, as already computed (\ref{fourthree}),
\be
{\cal L}_X \left( \mu(x)\, d^Dx\right) = \left( X^{\alpha}\, \mu_{,\alpha}
+ \mu\, {X^{\alpha}}_{,\alpha}\right)\, d^Dx
\ee
Therefore the basic equation (\ref{Divg}) is satisfied if, and only if
\be
\left( X^{\gamma}\, \mu_{,\gamma} + \mu\, {X^{\alpha}}_{,\alpha}\right)\,
\mu^{-1} = \frac{1}{2} \left( g^{\beta\alpha} X^{\gamma} 
g_{\alpha\beta , \gamma} + 2\, {X^{\alpha}}_{,\alpha}\right)\label{foureight}
\ee
i.e.
\bea
\frac{\mu_{,\gamma}}{\mu} & = & \frac{1}{2}\, g^{\beta\alpha} 
g_{\alpha\beta ,\gamma}\label{fivezero}\\
& = & \frac{1}{2}\, \partial_{\gamma} \ln |\det g\, |\\
\mu(x) & = & \mbox{ const. } |\det g(x)\, |^{1/2}.
\eea
The equation
\be
{\cal L}_X \omega_g = \frac{1}{2} {\rm Tr}\left(g^{-1} {\cal L}_X g\right)\,
\omega_g
\ee
has, up to multiplication by a constant, a unique solution
\be
\omega_g (x) = |\det g(x)|^{1/2} dx^1\wedge\ldots\wedge dx^D \; .
\qquad \blacksquare
\ee
We use the classical formula
\be
{\Gamma^{\alpha}}_{\alpha\gamma} = \frac{1}{2}\, g^{\beta\alpha} g_{\alpha
\beta ,\gamma}
\ee
with the Christoffel symbols ${\Gamma^{\alpha}}_{\beta\gamma}$; hence (\ref{foursix}) says
\be
\frac{1}{2} {\rm Tr} \left( g^{-1}\, {\cal L}_X g\right) = 
{X^{\alpha}}_{;\alpha}
= : {\rm Div}_g (X)
\ee
with the standard definition of the covariant divergence 
$X^{\alpha}_{\;\; ;\alpha}$ of the vector field $X$,
and we can write the basic equation (\ref{Divg}) in the form
\be
{\cal L}_X \omega_g = {\rm Div}_g(X) \cdot\omega_g \; .\label{fivesix}
\ee
If $X$ is a Killing vector field with respect to isometries, then
${\cal L}_X g = 0$, 
$\,{\cal L}_X \omega_g =0$ and $X_{\alpha\, ;\,\beta} + X_{\beta\, ;\,
 \alpha} =0$.
Hence ${X^{\alpha}}_{;\alpha} = 0$ and (\ref{fivesix}) is satisfied.

\subsection{The symplectic case $(M^D,\Omega)$, $D=2N$}

The symplectic form $\Omega$ on $M^{2N}$ is a closed 2-form of rank
$D=2N$. 
\be
\ba{rcll}
\Omega &=& \Omega_{AB} \, dx^A \wedge dx^B  & \mbox{ with }
A<B, \; \Omega_{AB} = -\Omega_{BA}, \; d\,\Omega=0  \\[.04in]
&=& {1\over 2} \, \Omega_{\alpha\beta} \, dx^{\alpha} \wedge dx^{\beta} &
\mbox{ no restriction on the order of $\alpha$, $\beta$} \\[.04in]
&=& {1\over 2} \, \Omega_{\alpha\beta} (
dx^{\alpha} \otimes dx^{\beta} -dx^{\beta} \otimes dx^{\alpha} )\\[.04in]
& = & \Omega_{\alpha\beta}\, dx^{\alpha}\otimes dx^{\beta}
\ea
\ee
since $\Omega_{\alpha\beta}=-\Omega_{\beta\alpha}$.\\
{\bf Remark}: There are two different definitions of the exterior
product, each with its concomitant definition of exterior derivative,
e.g.\ 
\bea
dx^1 \wedge dx^2 &=& dx^1 \otimes dx^2 - dx^2 \otimes dx^1 \\
\tilde{d}x^1 \tilde{\wedge}\, \tilde{d}x^2 &=& 
{1 \over 2}\left(\tilde{d}x^1 \otimes \tilde{d}x^2 -
\tilde{d}x^2 \otimes \tilde{d}x^1\right). 
\eea
With the second  definition, Stokes' formula for a $p$-form $\theta$ reads
$\int_M \tilde{d}\theta = 
(p+1)\int_{\partial M}\theta$;
with the first one it is simply $\int_M d\theta = \int_{\partial
M}\theta$. We choose the first definition, namely
\[
df^1\wedge\ldots\wedge df^p  =  \epsilon_{j_1\ldots j_p}\, df^{j_1}
\otimes\ldots\otimes df^{j_p}
\]
and in particular
\[
dx^1 \wedge \ldots \wedge dx^D   =  \epsilon_{j_1 \ldots j_D}
dx^{j_1} \otimes \ldots \otimes dx^{j_D}
\]
where $\epsilon$ is totally antisymmetric.
\ \\ \ \\
Since $\Omega$ is of rank $D=2N$, 
\[
\Omega^{\wedge N} := \Omega \wedge \ldots \wedge \Omega \qquad
\mbox{($N$ factors)}
\]
is a nonzero top form on $M^{2N}$ and the volume element
\bea
\omega_{\Omega} &=& {1 \over N!}\, \Omega^{\wedge N}  \label{volform} \\
&=& {\rm Pf}(\Omega_{\alpha\beta})\, d^D x = | \det 
\Omega_{\alpha\beta} \, |^{1/2}\, d^D x \; .
\eea
We shall show that the basic equation (\ref{DivO}) is satisfied if
and only if $\omega_{\Omega}$ is proportional to the volume form
 (\ref{volform}). \\
Proof: We define the inverse $\Omega^{-1}$ of $\Omega$,
calculate the quantity $\frac{1}{2} {\rm Tr} \left(\Omega^{-1} {\cal L}_X \Omega
\right)$, then prove the basic formula (\ref{DivO}).

\begin{itemize}
\item The symplectic form $\Omega$ defines an isomorphism from the 
tangent bundle $TM$ to the cotangent bundle $T^*M$ by
\[
\Omega : X \longmapsto i_X \Omega .
\]
We can then define
\be
X_{\alpha} := X^{\beta} \Omega_{\beta\alpha}
\ee
The inverse $\Omega^{-1} : T^*M\longrightarrow TM$ is given by
\[
X^{\alpha} = X_{\beta}\, \Omega^{\beta \alpha}
\]
with
\be
\Omega^{\alpha\beta}\, \Omega_{\beta\gamma} = \delta^{\alpha}_{\gamma} .
\ee
Note that in strict components, i.e. with $\Omega = \Omega_{AB}\, 
dx^A\wedge dx^B$ with $A<B$, $X_A$ is not equal to $X^B \Omega_{BA}$.
\item We compute
\bean
\left( {\cal L}_X \Omega\right)_{\alpha\beta} & = & X^{\gamma}\,
\Omega_{\alpha\beta,\gamma}+ \Omega_{\gamma\beta}\, X^{\gamma}_{\; ,\alpha}
+ \Omega_{\alpha\gamma}\, X^{\gamma}_{\; ,\beta}\\
 & = & X_{\beta ,\alpha} - X_{\alpha ,\beta}
\eean
using $d\Omega=0$, that is $\Omega_{\beta\gamma ,\alpha} + \Omega_{\gamma
\alpha ,\beta} + \Omega_{\alpha \beta, \gamma} =0$; hence
\[
{\left( \Omega^{-1} {\cal L}_X \Omega\right)^{\gamma}}_{\beta} =
\Omega^{\gamma \alpha} \left( X_{\beta ,\alpha} - X_{\alpha , \beta}\right)
\]
and
\be
\frac{1}{2} {\rm Tr} \left( \Omega^{-1} {\cal L}_X \Omega\right) = 
\Omega^{\gamma \alpha} X_{\gamma , \alpha} .\label{sevenfive}
\ee
\item According to Darboux' theorem,
there is a coordinate system $\left(x^{\alpha}\right)$ in
which the volume form $\omega_{\Omega} = \frac{1}{N!} \Omega^{\wedge N}$
is
\[
\omega_{\Omega} = dx^1 \wedge\ldots\wedge dx^{2N}.
\]
and $\Omega = \Omega_{\alpha\beta}\, dx^{\alpha} \otimes dx^{\beta}$ with 
{\it constant} coefficients $\Omega_{\alpha \beta}$; the inverse
matrix $\Omega^{\beta\alpha}$ is also made of constants, hence
${\Omega^{\beta\alpha}}_{,\gamma}=0$.\\
In these coordinates
\bea
{\cal L}_X \omega_{\Omega} & = & {X^{\alpha}}_{,\alpha}\, \omega_{\Omega}
\nonumber\\
& = & \left( X_{\beta}\, \Omega^{\beta\alpha}\right)_{,\alpha} 
\omega_{\Omega}\nonumber\\
& = & \left(X_{\beta ,\alpha}\, \Omega^{\beta\alpha} + X_{\beta}\, 
{\Omega^{\beta\alpha}}_{,\alpha}\right) \omega_{\Omega}\nonumber\\
& = & X_{\beta ,\alpha}\, \Omega^{\beta \alpha} \omega_{\Omega}
\eea
and we conclude by using (\ref{sevenfive}). $\qquad\blacksquare$
\end{itemize}
\ \\ \ \\
If $X$ is a Hamiltonian vector field, then ${\mathcal L}_X \Omega =0$
and ${\mathcal L}_X \omega_{\Omega} =0$.
The basic equation (\ref{DivO}) is trivially satisfied.

\section{Conclusion}
Integration by parts is the key to the progress made in this paper for
characterizing \index{volume forms} volume forms. It makes possible
 an infinitesimal
characterization of the translation invariant symbol ${\mathcal D}$,
\[
\int {\mathcal D}\varphi \, {\delta U \over \delta \varphi(x)} =0
\]
more powerful than its global translation (\ref{a8})
\[
{\mathcal D}(\varphi+\varphi_0)-{\mathcal D}\varphi =0 \; .
\]
The challenges we are now considering are the following:
\begin{itemize}
\item
Extending to infinite dimensional spaces the divergence formulae
(\ref{DivO}) and (\ref{Divg}).
\item
Clarifying the often observed relationship between the volume form
and the Schr\"{o}dinger equation satisfied by a functional integral.
\item
Developing issues mentioned briefly in this paper, in particular the 
Dobrushin-Lanford-Ruelle formula, and the annihilation/creation
operators defined by the \index{Malliavin formula} Malliavin formula. 
\item
Deriving the transformation laws of volume elements under the Cartan
development mapping between two spaces of pointed paths on different
manifolds. 
\item Extending the method from ordinary (bosonic) integration to Berezin
(fermionic) integration.
\end{itemize}

\section*{Acknowledgments}
Modern means of communication do not replace face-to-face
brainstorming, and a group based in Paris and Austin functions thanks
to contributions to travel expenses. We thank  a steadfast
friend John Tate who contributes to the annual visits of Pierre
Cartier to Austin, and we thank  the Jane and Roland Blumberg Centennial
Professorship for partial support of C\'ecile DeWitt-Morette's visits to
Paris. M. Berg is grateful to the Swedish Foundation for International
Cooperation in Research and Higher Education for financial support.

\end{document}